\documentclass[11pt, runningheads,a4paper]{llncs}
\usepackage{hyperref}
\usepackage[T1]{fontenc}
\usepackage{natbib}
\usepackage{amssymb}
\usepackage{placeins}
\usepackage[counterclockwise]{rotating}

\usepackage[left=3cm,right=3cm,top=3cm,bottom=3cm]{geometry}

\usepackage{graphicx}
\usepackage{float}

\usepackage{booktabs}      % \toprule \midrule \bottomrule
\usepackage{threeparttable}% notes under table (optional)
\usepackage{tabularx}      % auto-wrapping X columns
\usepackage{ragged2e}      % better ragged right
\usepackage{makecell}      % line breaks in cells 
\usepackage{amsmath}

\usepackage{color}

\begin{document}
%

% Title of the abstract
\title{A Statistical Framework for Detecting Emergent Narratives in Longitudinal Text Corpora}

% Author name(s)
\author{\textbf{Cynthia Medeiros}, John Quigley and Matthew Revie}

% Author affiliation and email
\institute{Department of Management Science, University of Strathclyde, UK \\
\email{cynthia.medeiros@strath.ac.uk}}

\maketitle              % 

%The abstract should not exceed 400 words. 
%The submission PDF, including references, must fit within a maximum of two pages.
%Abstract
\section*{Abstract}
Narratives about economic events and policies are widely recognised as influential drivers of economic and business behaviour. Yet the statistical identification of narrative emergence remains underdeveloped. Narratives evolve gradually, exhibit subtle shifts in content, and may exert influence disproportionate to their observable frequency, making it difficult to determine when observed changes reflect genuine structural shifts rather than routine variation in language use. 
\\~\\
We propose a statistical framework for detecting narrative emergence in longitudinal text corpora using Latent Dirichlet Allocation (LDA). We define emergence as a sustained increase in a topic’s relative prominence over time and articulate a statistical framework for interpreting such trajectories, recognising that topic proportions are latent, model-estimated quantities.
\\~\\
We illustrate the approach using a corpus of academic publications in economics spanning 1970-2018, where Nobel Prize-recognised contributions serve as externally observable signals of influential narratives. Topics associated with these contributions display sustained increases in estimated prevalence that coincide with periods of heightened citation activity and broader disciplinary recognition. These findings indicate that model-based topic trajectories can reflect identifiable shifts in economic discourse and provide a statistically grounded basis for analysing thematic change in longitudinal textual data.

\\

\section{Introduction}
\paragraph{}
A \textit{narrative} is understood as a structured account that organises events into a coherent sequence and assigns them meaning within a broader context, that is, a construction through which actors interpret and situate experience \citep{somers1994narrative}. At the individual level, narratives shape how agents interpret their environment, assess risks, and evaluate the plausibility of alternative courses of action.
Beyond this interpretative role, narratives may operate as dynamic social forces. As stories that circulate within public discourse, they can capture attention and shape collective understandings of economic and political events \citep{shiller2020narrative}. When particular narratives become prominent, they influence expectations and coordinate beliefs across individuals. Through this process, shifts in dominant narratives may translate into changes in financial, political, and societal decision-making, with aggregate consequences.
\paragraph{}
\cite{shiller2020narrative} argues that narratives evolve from individual discussions to collective prominence through a process of emergence, increase in traction, and eventual decline, often resembling a hump-shaped pattern. Once a narrative reaches its peak, it can be easily identified through the frequency with which it appears in discourse, and its decline can similarly be observed as references to it become less frequent over time. 
\paragraph{}
The early stages of emergence, however, are more difficult to observe. During this phase, they typically constitute only a small fraction of overall discourse, may be partially obscured by dominant stories, and often evolve through subtle shifts in emphasis or framing. As such, their identification requires formal methods capable of detecting structured change within discourse.
\paragraph{}
This article proposes a statistical framework for identifying the emergence of narratives in longitudinal data. Since narratives are expressed and transmitted through language, we treat text as observable data from which latent thematic structure can be inferred. We employ Latent Dirichlet Allocation (LDA), a probabilistic topic model, to recover the underlying themes (`topics') in a collection of documents \citep{blei2003latent}, and to estimate their relative prominence over time. 
\paragraph{}
A range of extensions to LDA has been developed, including the Dynamic Topic Model \citep{blei2006dynamic}, which allows topic–word distributions to evolve over time; the Structural Topic Model \citep{roberts2013structural}, which incorporates document-level metadata; and more recent neural or embedding-based approaches, such as the BERTopic Model \citep{grootendorst2022bertopic}. These frameworks are often motivated by predictive performance or representational richness. However, they also introduce additional layers of modelling and tuning choices that can complicate interpretation when the primary object of interest is a statistically interpretable measure of thematic prominence over time. The present study therefore adopts baseline LDA deliberately. Our objective is not to maximise predictive accuracy, but to examine what can be inferred from the posterior distribution of a generative model about the structure of discourse over time, a central research stream in topic modelling identified by \cite{blei2012probabilistic}. LDA is particularly well suited to this goal because it yields an explicit probabilistic representation of document-level topic proportions that can be aggregated across time periods to quantify changes in thematic prominence within discourse. Moreover, fixing topics over the sample avoids the comparability issues that may arise when topic definitions themselves evolve, thereby providing a stable reference frame for interpreting temporal change. Recent work on emergence detection also suggests that probabilistic topic models such as LDA can provide reliable signals for retrospective identification of emerging themes, in some settings outperforming embedding-based alternatives such as BERTopic \citep{li2025evaluation}. Embedding-based approaches remain valuable when the objective is to identify semantically novel concepts that may not yet have stable lexical realisations \citep{ma2025identifying}, and hybrid pipelines combining topic models with large language models (LLMs) have been proposed to detect and interpret narrative shifts \citep{lange2025narrative}. Our approach is complementary. The contribution of this article lies not in proposing a new topic model, but in developing a statistical framework for interpreting changes in estimated topic prevalence as signals of narrative emergence. LDA serves as a probabilistic measurement model that produces coherent estimates of thematic prominence, whose temporal dynamics can then be evaluated against external indicators of influence.

\paragraph{}
This perspective aligns naturally with \citet{shiller2020narrative}'s account of narratives as evolving thematic constructions that gain prominence within discourse. If narratives manifest as structured re-allocations of thematic attention, then posterior estimates of topic proportions provide a principled statistical object through which such re-allocations can be measured. By focusing on LDA, we isolate the inferential properties of a well-understood generative model and clarify how sustained changes in its posterior summaries may be interpreted as evidence of narrative emergence.
\paragraph{}
The main goals of this article are threefold. First, to formalise narrative emergence as a sustained increase in estimated topic proportions within longitudinal text data. Second, to clarify how changes in these estimated topic proportions can be interpreted as evidence of narrative formation. Third, to assess whether shifts in topic proportions are associated with measurable indicators of influence. In Section \ref{section2}, we present the statistical framework and define narrative emergence in terms of estimated topic proportions. Section \ref{section3} describes the data and empirical strategy. In Section \ref{section4}, we report the empirical results, examining the temporal dynamics of topic proportions and their relationship with indicators of influence. Section \ref{section5} concludes with a discussion of interpretation, limitations, and broader implications.

\section{Statistical Framework for Narrative Emergence}
\label{section2}
\paragraph{}
LDA, introduced by \cite{blei2003latent}, builds upon earlier mixture models for text, most notably probabilistic Latent Semantic Analysis (pLSA) \citep{hofmann1999probabilistic}. In pLSA, each document is represented as a mixture of latent topics, but the topic proportions for each document are estimated independently as free parameters. As the number of documents increases, the number of such parameters grows accordingly, which may lead to overfitting and limits the model's ability to generalise to new documents. LDA addresses this limitation by introducing a Dirichlet prior over document-level topic proportions, inducing a shared probabilistic structure across the corpus (collection of documents). Under LDA, each document is therefore modelled as a mixture of latent topics, where a topic is defined as a probability distribution over a fixed vocabulary (a set of unique words present in the corpus), and documents differ in the proportions with which they express these topics.
\paragraph{}
Let $D$ denote the number of documents in the corpus and $K$ the number of topics. For document $d$, let $\theta_{d}=(\theta_{d1},\cdots,\theta_{dK})$ denote the vector of topic proportions, where $\theta_{dk}$ represents the share of document $d$ devoted to topic $k$. Each topic $k$ is associated with a word distribution $\beta_{k}$, a probability vector over the vocabulary. The generative structure of LDA assumes that topic proportions $\theta_{d}$ are drawn from a Dirichlet distribution with parameter $\alpha$, and topic-specific word distributions $\beta_{k}$ are drawn from a Dirichlet distribution with parameter $\eta$. Conditional on $\theta_{d}$, each word in document $d$ is generated by first sampling a topic assignment and then sampling a word from the corresponding topic distribution. More formally, the generative process can be described as \citep{blei2003latent}:
\begin{enumerate}
  \item For each topic $k = 1,\cdots,K$:
  \begin{itemize}
    \item Draw the per-topic word distribution
    \[
      \beta_k \sim \mathrm{Dirichlet}(\eta).
    \]
  \end{itemize}

  \item For each document $i = 1,\cdots,D$:
  \begin{itemize}
    \item Draw the topic-proportion vector
    \[
      \theta_d \sim \mathrm{Dirichlet}(\alpha).
    \]
    \item For each word $n = 1,\cdots,N_i$:
    \begin{enumerate}
      \item[(a)] Draw the topic assignment
      \[
        z_{dn} \sim \mathrm{Multinomial}(\theta_d).
      \]
      \item[(b)] Draw the observed word
      \[
        w_{dn} \sim \mathrm{Multinomial}(\beta_{z_{dn}}).
      \]
    \end{enumerate}
  \end{itemize}
\end{enumerate}

\paragraph{}
Formally, the joint distribution of the latent and observed variables can be written as 
\begin{equation}
    p(\beta, \theta, z, w)=\prod_{k=1}^{K}p(\beta_{k} | \eta) \prod_{d=1}^{D} p(\theta_{d}|\alpha) \prod_{n=1}^{N} p(z_{dn} | \theta_{d}) p(w_{dn} | \beta_{z_{dn}})
\end{equation}
where $z_{dn}$ denotes the latent topic assignment for word $n$ in document $d$, and $w_{dn}$ the observed word.
\paragraph{}
Exact posterior inference under this model is intractable \citep{blei2003latent}. In practice, approximate methods such as variational inference \citep{blei2003latent} or Gibbs sampling \citep{steyvers2007probabilistic} are used to obtain posterior summaries. In this study, we use the posterior mean of the document-level topic proportions $\hat{\theta}_{dk}$ as our quantity of interest. These posterior mean estimates represent the model-based share of document $d$ devoted to topic $k$ and serve as summaries of thematic composition.
\paragraph{}
When documents are indexed by time, the estimated topic proportions can be aggregated within time periods to construct a time series of thematic composition. Let $t=1, \cdots,T$ index discrete time periods, and let $\mathcal{D}_{t}$ denote the set of documents observed in period $t$. For each topic $k$, we define the average topic proportion in period $t$ as

\begin{equation}
\label{eq:bartheta}
    \bar{\theta}_{k,t} = \frac{1}{|\mathcal{D}_{t}|} \sum_{d \in \mathcal{D}_{t}} \hat{\theta}_{dk}
\end{equation}

$\bar{\theta}_{k,t}$ represents the estimated share of discourse in period $t$ associated with topic $k$. It provides a comparable measure of thematic prominence across time: since all documents are modelled as mixtures of the same set of topics, changes in $\bar{\theta}_{k,t}$ reflect shifts in the relative emphasis placed on recurring thematic patterns within the corpus.
\paragraph{}
For each topic $k$, the aggregated quantity $\bar{\theta}_{k,t}$ defines a time series of estimated topic proportions. We interpret narrative emergence as a persistent upward shift in this series relative to its earlier level. In practical terms, emergence occurs when a theme that previously occupied a small share of discourse exhibits a sustained increase in its estimated topic proportion, corresponding to the rise-peak-decline dynamic described by \cite{shiller2020narrative}. 

\section{Data and Empirical Strategy}
\label{section3}
\paragraph{}
To illustrate the proposed framework, we examine the evolution of discourse within academic economics. Journal publications provide a record of disciplinary debate spanning several decades, making it possible to follow how thematic concerns develop over time. Unlike public or media discussions, which may shift quickly in response to immediate events, academic research typically evolves more gradually: new lines of research are introduced through foundational contributions, adopted by other scholars, debated and refined, and eventually incorporated into mainstream research. As this process unfolds, it reshapes the distribution of attention within the literature. Early contributions may occupy a relatively small share of discussion; over time, related terminology becomes more common, associated questions expand, and the research agenda broadens. The resulting transformation is not a single break, but a sustained reallocation of thematic emphasis across the corpus. 
\paragraph{}
Economics offers a particularly structured setting for examining these dynamics. The discipline is organised around a well-established classification system, JEL codes, that delineates subfields and provides a consistent framework for categorising research over time. Economic research is highly heterogeneous, spanning theoretical, empirical, and methodological traditions that evolve at different rates, and often respond to distinct intellectual and policy influences \citep{hamermesh2013six}. As such, innovations that reshape discourse in one subfield may not translate directly into others. Analysing subfield-specific corpora defined by JEL codes therefore provides a more coherent thematic context, reducing cross-domain heterogeneity and improving the interpretability of temporal changes in discourse. 
\paragraph{}
Additionally, economics exhibits a highly centralised publication hierarchy. A small number of general-interest journals -- the `Top-5': American Economic Review (AER), Quarterly Journal of Economics (QJE), Journal of Political Economy (JPE), Econometrica (ECMA), Review of Economic Studies (REStud) -- play a disproportionate role in shaping mainstream research agendas \citep{ellison2002slowdown, card2013nine, heckman2020publishing}. As a result, shifts in thematic emphasis within these outlets tend to reflect broader changes in disciplinary attention rather than localised debates. Intellectual developments are therefore traceable both within subfields and at the core of the discipline. Finally, the discipline also provides identifiable points at which particular lines of work receive broad professional recognition. The awarding of the Nobel Prize in Economic Sciences acknowledges contributions that have substantially influenced the direction of the discipline, marking a stage at which that body of work has become widely established. This makes it possible to examine whether thematic prominence increases in the period preceding such recognition. Figure \ref{fig:emergence_schematic} illustrates this process.

\begin{figure}[t]
    \centering
    \includegraphics[width=0.65\textwidth]{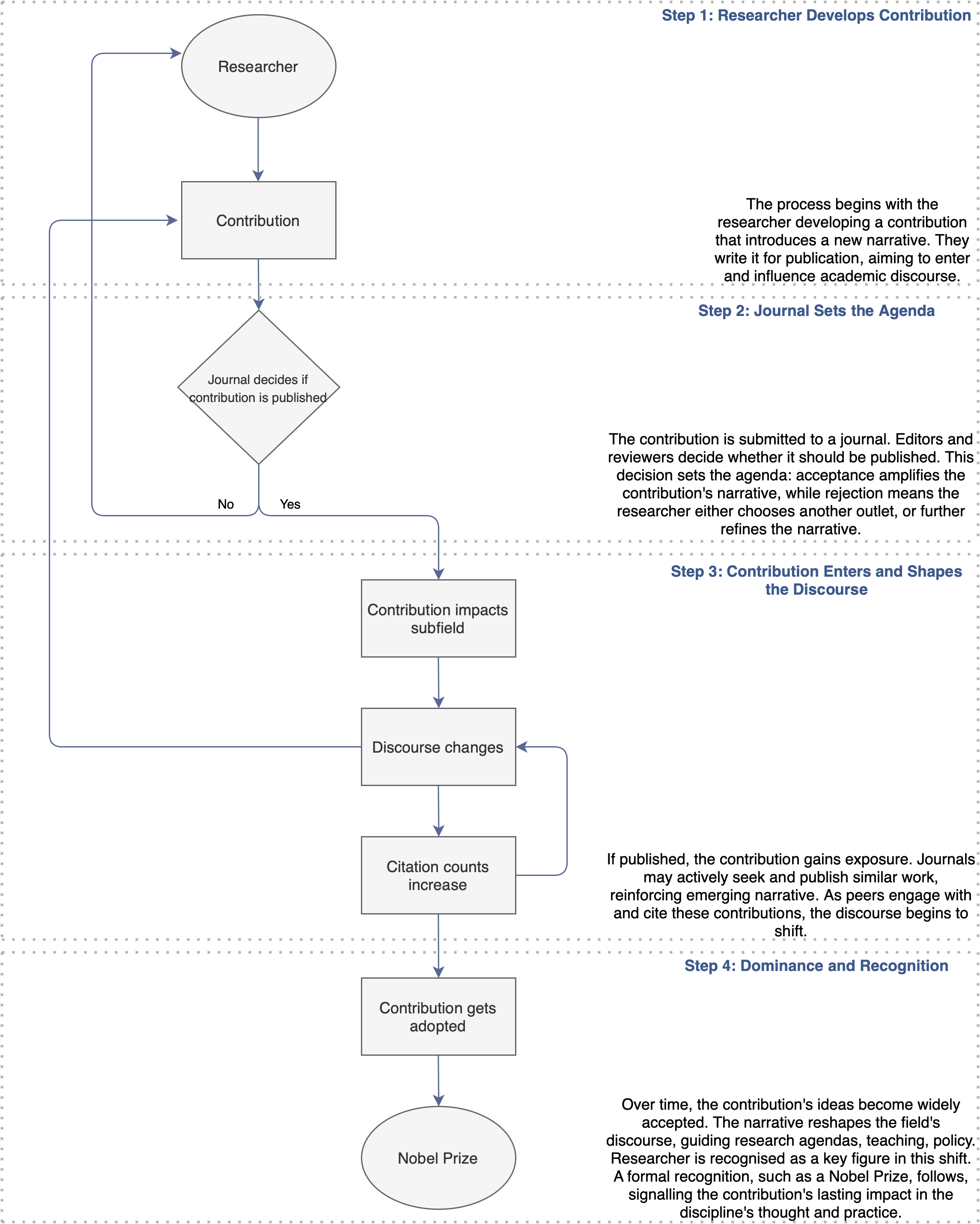}
    \caption{Flowchart illustrating the stages of a Narrative Shift in academic thought. In our case study, we are mainly concerned with using NLP to identify the changes occurred in Step 3.}
    \label{fig:emergence_schematic}
\end{figure}
\FloatBarrier
\paragraph{}
Full-text articles were obtained from JSTOR's Data for Research platform (Constellate), which provided access to digitised journal archives and associated metadata. The corpus consists of 17,877 research articles published in AER, QJE and JPE, which are consistently identified as the most influential among the `Top-5' general-interest journals. \citep{card2013nine, hamermesh2013six}. The sample period begins in 1970, when JEL classification was introduced across the field, allowing discourse to be analysed within a consistent subfield structure, and concludes in 2018, the last complete publication year available at the JSTOR database at the time of data extraction. From this archive, subfield-specific corpora were constructed using JEL classification codes, and topic models were estimated separately within each subfield.
\paragraph{}
The full text of each article, indexed by the publication year, was standardised using conventional natural language pre-processing procedures. The JSTOR archive provided tokenised text; from this, non-alphabetic characters, numerical expressions, and stopwords (e.g. `the', `and', `of') were removed. Generic academic terms with limited semantic content were excluded to reduce noise. All tokens were lowercased. Stemming or lemmatisation was not applied, in order to preserve distinctions between conceptually meaningful variants (for example, `institution' and `institutional').

\paragraph{}
After preprocessing, the corpus was transformed into a document-term matrix, where each row corresponds to a document and each column to a term in the resulting vocabulary. Matrix entries record within-document word frequencies. This representation defines the observed data on which LDA is estimated, yielding document-specific topic proportions that can subsequently be aggregated by publication year.

\paragraph{}
To examine whether thematic prominence increases prior to formal professional recognition, we focus on Nobel Prize-recognised contributions introduced in or after 1980. This restriction reflects both empirical and design considerations. Empirical evidence indicates a lag between initial publication of a contribution and its subsequent Nobel recognition in economics, estimated at approximately $23 \pm 7.5$ years \citep{bjork2020journals}. Given that the corpus begins in 1970, selecting 1980 as the earliest introduction year ensures that a meaningful pre-recognition period is observable within the data for each contribution. In particular, contributions introduced in the 1980s and 1990s are typically recognised in the 2010s and 2020s, allowing the analysis to capture both the diffusion phase preceding formal recognition and the consolidation phase that follows. This restriction therefore ensures that narrative trajectories can be examined over a sufficiently long temporal horizon within the 1970-2018 window.
\paragraph{}
In our analysis, we exclude awards primarily associated with econometric methodology, as their diffusion reflects technical developments rather than shifts in substantive research discourse. The selected contributions span major subfields defined by JEL classification, enabling narrative trajectories to be examined within relatively coherent thematic domains. A complete list of included laureates and associated publications is provided in the Appendix.
\paragraph{}
To complement the analysis of topic trajectories, we compare the annual evolution of $\bar{\theta}_{k,t}$ with citation counts for the corresponding contributions. Citation counts are a conventional bibliometric indicator of scholarly influence and provide an observable benchmark against which changes in thematic prominence can be interpreted \citep{hamermesh2013six, card2013nine}.

\section{Empirical Results}
\label{section4}
\subsection{Topic Identification}
\label{sec:topicident}
\paragraph{}
We begin by examining whether the LDA models recover thematically coherent structures that align with Nobel Prize-recognised contributions introduced after 1980. Within each subfield-specific corpus, the number of topics $K$ was selected through iterative estimation. While likelihood-based criteria such as perplexity assess statistical fit, they do not guarantee interpretability of topic-word distributions; prior work has shown that human evaluation more reliably captures thematic coherence in topic modelling applications \citep{chang2009reading}. We therefore retained specifications that yielded clearly distinguishable and interpretable clusters of terms.
\paragraph{}
Topic identification proceeds by examining the highest-probability words within each topic given the $\beta_{k}$ posterior and evaluating whether the resulting lexical clusters correspond to conceptual or methodological orientation of the recognised contributions. A topic is considered aligned with a Nobel-recognised contribution only if its dominant vocabulary reflects a coherent research strand rather than isolated keywords. Table \ref{tab:nobel_topics_lda}  summarises the topics identified as most closely associated with the selected laureates.
\paragraph{}
Results indicate that the model recovers several Nobel-associated research strands as lexically distinct topics, but not all. In \textit{development economics}, one topic is characterised by terms such as `treatment', `experiment', `school', `child', `health', and `intervention'. The concentration of terminology related to field experimentation and programme evaluation reflects the empirical paradigm associated with the introduction of randomised controlled trials by laureates Abhijit Banerjee, Esther Duflo and Michael Kremer \citep{miguel2004worms, duflo2007using, banerjee2015miracle}. A separate development-related topic includes terms such as `institution', `democracy', `government', and `property', corresponding to research emphasising the long-run role of political institutions in shaping economic outcomes, introduced by Daron Acemoglu, Simon Johnson and James Robinson \citep{acemoglu2001colonial, acemoglu2005institutions, robinson2012nations}. The lexical separation of these clusters suggests that the model distinguishes between alternative intellectual approaches within the same subfield: one centred on methodological innovation and micro-empirical evaluation, the other on theoretical and historical analysis.
\paragraph{}
In \textit{macroeconomic research}, a further topic associated with financial crises and banking instability is identified, characterised by vocabulary related to credit markets, banking systems, liquidity constraints, and crisis transmission. This cluster aligns with the literature surrounding financial intermediation and systemic risk, including contributions by Ben Bernanke \citep{bernanke2001should} and  Douglas Diamond and Philip Dybvig \cite{diamond1983bank, diamond1984financial}. Unlike foundational asset-pricing terminology, which is pervasive within financial economics, this topic reflects concentrated discourse around banking fragility and crisis dynamics, indicating the presence of a coherent research strand.
\paragraph{}
Within \textit{labour economics}, the model identifies a distinct topic centred on gender and labour market dynamics, characterised by terms such as female, wage, participation, education, and gender. The concentration of this vocabulary distinguishes it from broader human capital or earnings-related clusters. The topic aligns closely with research on female labour force participation and gender wage inequality by Claudia Goldin \citep{goldin1990understanding, goldin2006quiet, goldin2014grand}.
\paragraph{}
By contrast, the contributions of Paul Romer, David Card, and Robert Shiller do not correspond clearly to distinguishable topics recovered by LDA. In Romer's case, his contribution on endogenous growth theory \citep{romer1986increasing, romer1990endogenous} extends and reformulates the neoclassical growth theory pioneered by \cite{solow1957technical}, drawing heavily on an already established macroeconomic vocabulary. As a result, the associated discourse remains embedded within dominant growth-related themes rather than appearing as a separate lexical cluster.
\paragraph{}
Similarly, Card's contributions to labour economics are primarily methodological advancing quasi-experimental approaches to causal inference within applied microeconomics \citep{card1993minimum, card2001immigrant}. These methods are applied across diverse contexts and do not rely on a narrow subject-matter lexicon. As a consequence, the language associated with this research appears dispersed across broader applied topics, rather than within a distinct thematic grouping.
\paragraph{}
Shiller's recognised work spans multiple domains within financial economics, including asset return predictability, behavioural finance, and housing market measurement \citep{shiller1990market, shiller1990speculative, shiller2003efficient}. The dispersion of these contributions across related but distinct areas makes it difficult for the model to associate them with a single coherent vocabulary. In this case, the absence of a dedicated topic reflects the distributed lexical expression of the work within the corpus.
\paragraph{}
Taken together, these cases illustrate that LDA can capture contributions as distinct topics when they are expressed through concentrated and recognisable thematic vocabularies. Where influential research extends existing paradigms, operates methodologically across fields, or spans multiple substantive domains, its influence may be reflected in shifts within broader topics rather than through the formation of a separate cluster.

\begin{table}[t]
\centering
\begin{threeparttable}
\caption{Nobel-associated topics identified by LDA}
\label{tab:nobel_topics_lda}
\small
\setlength{\tabcolsep}{4pt}
\renewcommand{\arraystretch}{1.15}

\begin{tabularx}{\textwidth}{@{} l l >{\RaggedRight\arraybackslash}X >{\RaggedRight\arraybackslash}X l @{}}
\toprule
\textbf{Subfield} & \textbf{Topic ID} & \textbf{Representative Words} & \textbf{Interpreted theme} & \textbf{Associated laureate(s)} \\
\midrule
Development & T8 &
experiment, treatment, design, school, health  &
Experimental development / RCT &
Banerjee, Duflo, Kremer \\

Development & T9 &
institution, democracy, government, electoral, vote &
Institutional political economy &
Acemoglu, Johnson, Robinson \\

Labour & T7 &
female, wage, labor, education, school  &
Gender and labour markets &
Goldin \\

Macroeconomics & T6 &
effect, control, coefficient, design, treatment &
Experimental development / RCTs &
Banerjee, Duflo, Kremer \\

Macroeconomics & T12 &
bank, asset, risk, borrow, credit  &
Asset pricing / market dynamics &
Ben Bernanke, Diamond \& Dybvig \\

Finance & T4 &
vote, election, control, institution, government &Institutional political economy &
Acemoglu, Johnson, Robinson \\

Finance & T7 &
dynamic, aggregate, shock, investment, borrow &
Financial intermediation and macro-financial shocks &
Ben Bernanke, Diamond \& Dybvig \\

\bottomrule
\end{tabularx}

\begin{tablenotes}[flushleft]
\footnotesize
\item \textit{Notes:} Top terms shown are the ten highest-probability words within each topic; ellipses indicate omitted terms.
\end{tablenotes}
\end{threeparttable}
\end{table}
\FloatBarrier 

\subsection{Temporal Dynamics and Trend Detection}
\label{sec:tempdynamics}
\paragraph{}
To examine whether the trajectories of the topics in Section \ref{sec:topicident} exhibit sustained temporal change consistent with narrative emergence, we analyse the evolution of the aggregated topic proportions $\bar{\theta}_{k,t}$ defined in Equation \ref{eq:bartheta}. Figure \ref{fig:emergence_schematic} presents the resulting time series for seven selected topics associated with the Nobel-recognised research strands across finance, macroeconomics, labour, and growth, between 1970 and 2018. Visual inspection suggests heterogeneous patterns of increase across topics. Finance - Topic 7 and Macroeconomics - Topic 12 display particularly pronounced upward trajectories, while the remaining topics exhibit more gradual but consistently rising trajectories. 

\begin{figure}[t]
    \centering
    \includegraphics[width=\textwidth]{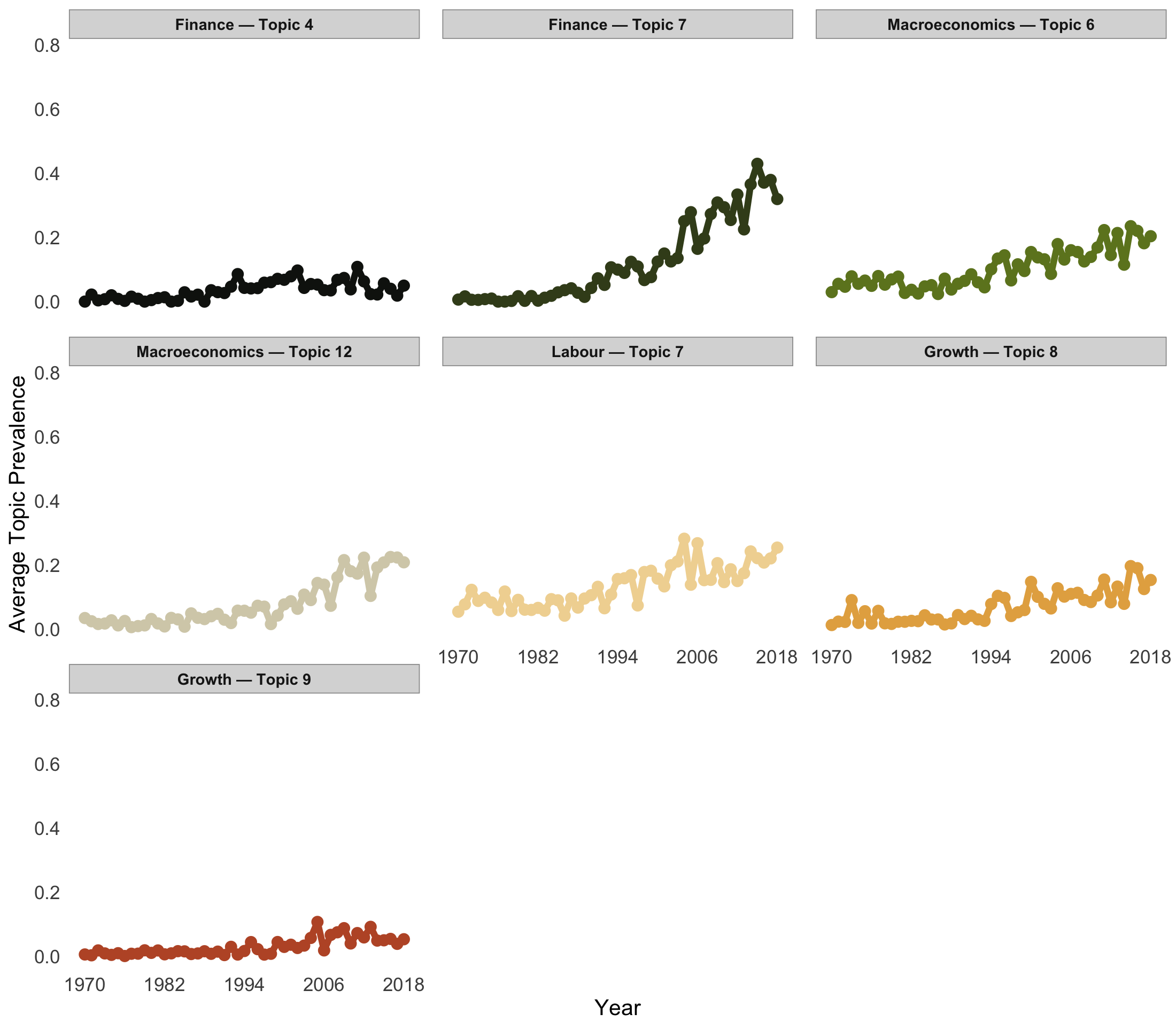}
    \caption{Time series of $\bar{\theta}_{k,t}$ for selected Nobel-associated topics across subfields, 1970-2018. The series indicate sustained increases in Finance-Topic 7 and Macroeconomics-Topic 12, alongside more moderate upward movements in the other topics.}
    \label{fig:emergence_schematic}
\end{figure}
\paragraph{}
Consistent with the definition of narrative emergence as a sustained increase in $\bar{\theta}_{k,t}$, we use the Mann-Kendall test to evaluate whether each trajectory exhibits a monotonic upward trend over time. Alongside statistical significance, we report Kendall's $\tau$ as a measure of the strength of monotonic association between time and topic prevalence. To characterise the magnitude of change, we complement the test with Sen's slope estimator, which provides a robust estimate of the median annual rate of increase together with a confidence interval. 
\paragraph{}
The results of the trend analysis are reported in Table \ref{tab:mk_trends}. All seven trajectories exhibit statistically significant positive trends over the study period. Kendall's $\tau$ values range from 0.47 to 0.84, indicating moderate to very strong monotonic increases in topic prominence. The strongest trend is observed for Finance - Topic 7 $(\tau = 0.835)$, which also displays the largest estimated annual increase in prevalence (Sen's slope = 0.0076, 95\% CI [0.0061, 0.0091]). Macroeconomics - Topic 12 similarly exhibits a strong upward trajectory $(\tau = 0.721)$, with a median annual increase of approximately 0.0038. The remaining topics show smaller but consistently positive slopes, with confidence intervals excluding zero in all cases, indicating that the estimated rates of increase are statistically distinguishable from zero.
\paragraph{}
The magnitude of these slope estimates provides additional substantive insight into the pace of thematic change. For example, the estimated growth rate for Finance - Topic 7 implies an increase of approximately 0.3 in average topic prevalence in over four decades, consistent with the rise observed in Figure \ref{fig:emergence_schematic}. Macroeconomics - Topic 12, which exhibits similarly strong growth, is associated also with closely related research on financial intermediation and macro-financial instability, linked to the contributions of Bernanke and Diamond and Dybvig. The prominence of these trajectories is  consistent with the increasing attention to financial crises and systemic risk following the global financial crisis of 2007-2009. In contrast, the growth trajectories associated with Topics 8 and 9 in the growth subfield are more gradual, reflecting slower consolidation of thematic prominence. These differences suggest heterogeneity in the speed with which research narratives gain prominence across subfields, consistent with variation in intellectual diffusion and adoption dynamics across areas of economics.
\\~\\
\begin{table}[t]
\centering
\caption{Mann--Kendall trend tests for annual aggregated topic prevalence (1970--2018) for selected Nobel-associated topics across subfields. Kendall’s $\tau$ measures the strength of the monotonic trend, and Sen’s slope estimates the median annual increase in topic prevalence. All trajectories exhibit statistically significant positive trends.}
\label{tab:mk_trends}
\small
\setlength{\tabcolsep}{6pt}
\begin{tabular}{lcccc}
\toprule
\textbf{Subfield -- Topic} & $\tau$ & \textbf{p-value} & \textbf{Sen's slope} & \textbf{95\% CI} \\
\midrule
Finance -- Topic 7            & 0.835 & $< 10^{-6}$          & 0.0076 & [0.0061, 0.0091] \\
Macroeconomics -- Topic 6     & 0.621 & $< 10^{-6}$          & 0.0035 & [0.0027, 0.0043] \\
Macroeconomics -- Topic 12    & 0.721 & $< 10^{-6}$          & 0.0038 & [0.0029, 0.0047] \\
Labour -- Topic 7             & 0.607 & $< 10^{-6}$          & 0.0034 & [0.0028, 0.0041] \\
Growth -- Topic 8             & 0.628 & $< 10^{-6}$          & 0.0025 & [0.0019, 0.0032] \\
Growth -- Topic 9             & 0.558 & $< 10^{-6}$          & 0.0011 & [0.0009, 0.0014] \\
Finance -- Topic 4            & 0.469 & $2.0 \times 10^{-6}$  & 0.0013 & [0.0008, 0.0018] \\
\bottomrule
\end{tabular}
\end{table}
\FloatBarrier

\subsection{Relationship with citations}
\label{sec:citations}
\begin{figure}[t]
    \centering
    \includegraphics[width=\textwidth]{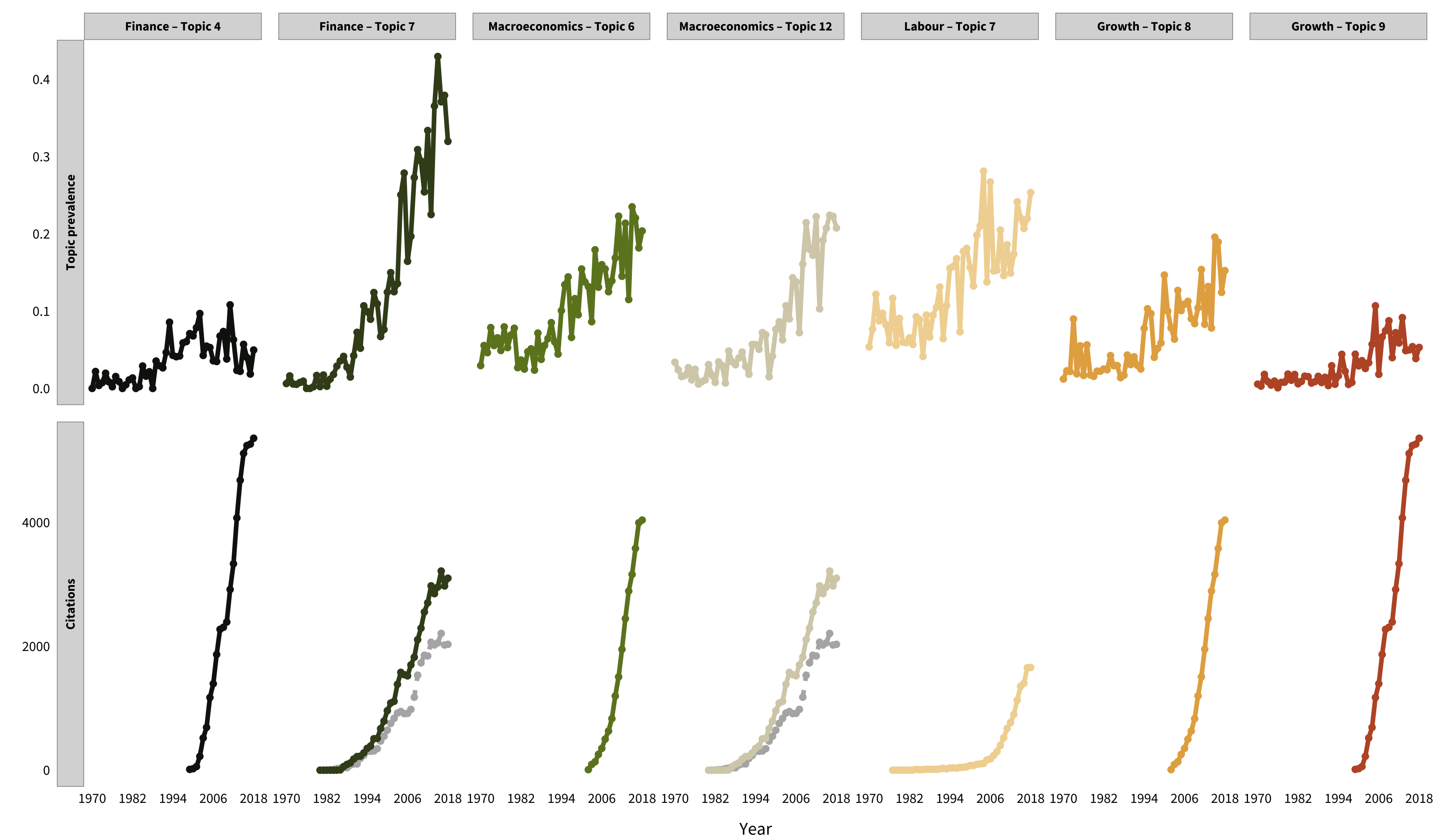}

  %  \vspace{2mm}
 %   \includegraphics[width=0.55\textwidth]{legend_topics_citations.pdf}

    \caption{
    Topic prevalence and citation trajectories for selected Nobel-associated topics, 1970-2018. Upper panels show annual aggregated topic prevalence $\bar{\theta}_{k,t}$; lower panels show citation counts for the corresponding Nobel-recognised contributions. For jointly associated topic-contribution pairs, Bernanke citations are shown in grey and Diamond-Dybvig citations in the topic colour. The panels illustrate both contemporaneous alignment (e.g. Finance Topic 7) and apparent lead–lag patterns (e.g. Labour Topic 7), motivating the formal lag analysis.
    }
    \label{fig:prevalence_citations}
\end{figure}

\paragraph{}
To examine whether statistically detected narrative emergence corresponds to measurable scholarly influence, we analyse the temporal association between annual aggregated topic prevalence $\bar{\theta}_{k,t}$ and citation counts for the corresponding Nobel-recognised contributions.
\paragraph{}
Figure \ref{fig:prevalence_citations} presents the joint trajectories of topic prevalence and citation counts for the selected topics. In several cases -- notably Finance - Topic 7 and Macroeconomics - Topic 12 -- the two series appear to move closely together, with pronounced increases in topic prevalence occurring alongside sharp increases in citations. In other cases, such as Labour - Topic 7 and Growth - Topic 9, the increase in thematic prominence appears to precede the rapid acceleration of citations. These visual patterns suggest heterogeneous temporal relationships across subfields and motivate a more formal lead-lag analysis.

\FloatBarrier

\begin{figure}[t]
    \centering
    \includegraphics[width=\textwidth]{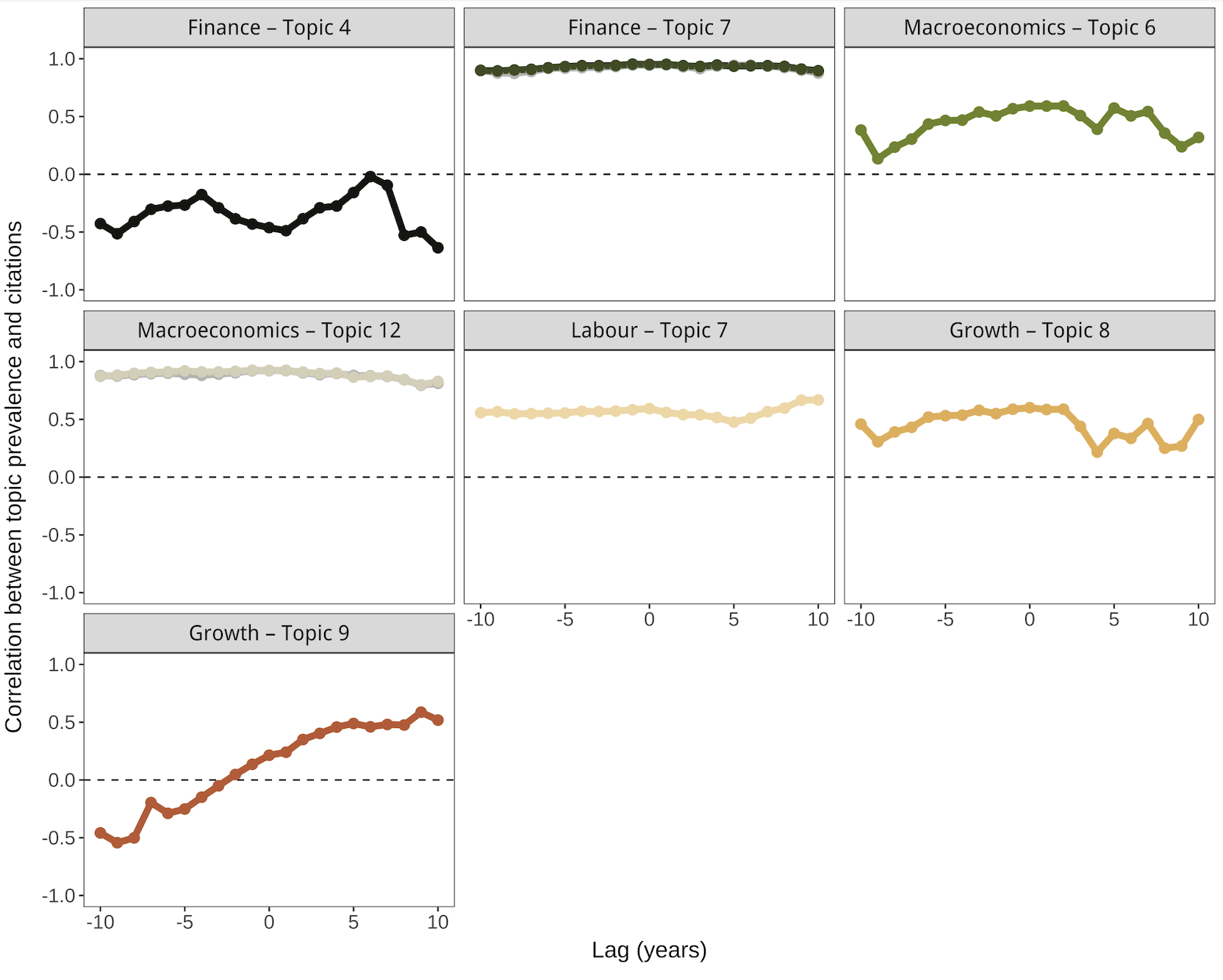}

    \caption{
   Lagged correlations between annual topic prevalence $\bar{\theta}_{k,t}$ and citation counts for selected Nobel-associated topics. The horizontal axis denotes lag in years; positive values indicate that topic prevalence precedes citation growth. Finance–Topic 7 and Macroeconomics–Topic 12 exhibit high correlations near lag zero, while Growth–Topic 9 displays increasing correlation at positive lags, consistent with topic prevalence leading subsequent citation accumulation.
    }
    \label{fig:lag_correl}
\end{figure}

\paragraph{}
Let $C_{t}$ denote the annual citation count for a given contribution. For each topic-contribution pair, we compute lagged correlations between $\bar{\theta}_{k,t}$ and $C_{t+\ell}$ for lags $\ell \in [-10, 10]$. The symmetric ten-year window is chosen to capture medium-run diffusion dynamics within the observable sample. Empirical evidence indicates that recognition in economics unfolds over extended horizons \citep{bjork2020journals}. Positive values of $\ell$ indicate that increases in topic prevalence precede citation growth; negative values indicate the reverse. The optimal lag is defined as 

\begin{equation}
    \ell^{*} = \arg \max_{\ell} \text{corr} (\bar{\theta}_{k,t}, C_{t+\ell})
\end{equation}
\paragraph{}
Figure \ref{fig:lag_correl} provides the full lag-correlation profiles for each topic-contribution pair. Table \ref{tab:lag_correlations} summarises the maximum correlation, the corresponding lag, and the contemporaneous correlation for each case. Three distinct temporal structures emerge.
\paragraph{}
First, several topics exhibit very high correlations at or near lag zero. Finance - Topic 7 and Macroeconomics - Topic 12 both display correlations exceeding 0.92 in absolute value, with peak alignment occurring either contemporaneously or within one year. In these cases, thematic prominence and citation accumulation evolve in close synchrony. Second, Labour - Topic 7 and Growth - Topic 9 exhibit peak correlations at positive lags of approximately 9-10 years. In these cases, the contemporaneous correlations are materially lower than the maximum lagged correlations. For example, Growth - Topic 9 displays a modest correlation at lag zero, but a substantially stronger association at longer positive lags. This pattern indicates that increases in thematic prominence precede citation acceleration, suggesting a gradual consolidation process in which a research strand first expands with discourse before being reflected in formal citation metrics. Third, Finance - Topic 4 displays weaker and negative correlations across lags, despite exhibiting a statistically significant upward trend in Section \ref{sec:tempdynamics}. The absence of strong citation alignment in this case is informative. Not all sustained increases in topic prevalence correspond to concentrated citation dynamics. Thematic expansion may reflect broader integration into multiple research strands rather than diffusion centred on a small number of highly cited contributions.
\paragraph{}
Taken together, the lag analysis strengthens the interpretation of narrative emergence as more than lexical drift. In most cases, statistically significant increases in topic prevalence align with, or precede, measurable shifts in scholarly recognition. Moreover, the variation in lag structure suggests that narrative diffusion is not uniform across subfields. Some narratives rise in tandem with citation growth, while others exhibit a lead-lag structure consistent with delayed institutional consolidation.
\paragraph{}
The framework therefore provides two complementary layers of inference: (i) internal statistical evidence of sustained thematic change, (ii) external validation through alignment with citation trajectories.

\FloatBarrier

\begin{sidewaystable*}[p]
\centering
\begin{threeparttable}
\caption{Peak lag correlations between annual topic prevalence $\bar{\theta}_{k,t}$ and citation counts $C_t$. The best lag $\ell^*$ maximises $\mathrm{corr}(\bar{\theta}_{k,t}, C_{t+\ell})$ over $\ell \in [-10,10]$.}
\label{tab:lag_correlations}
\small
\begin{tabular}{l l c c c l}
\toprule
\textbf{Subfield – Topic} & \textbf{Laureate(s)} & \textbf{Best Lag} & \textbf{Max Corr.} & \textbf{Corr. (Lag 0)} & \textbf{Pattern} \\
\midrule
Finance – Topic 7 & Bernanke & 1  & 0.947 & 0.944 & Near-contemporaneous \\
Finance – Topic 7 & Diamond \& Dybvig & -1 & 0.954 & 0.951 & Near-contemporaneous \\
Macroeconomics – Topic 12 & Bernanke & 0  & 0.922 & 0.922 & Contemporaneous \\
Macroeconomics – Topic 12 & Diamond \& Dybvig & -1 & 0.925 & 0.923 & Near-contemporaneous \\
Macroeconomics – Topic 6 & Banerjee, Duflo \& Kremer & 2  & 0.591 & 0.590 & Slight topic lead \\
Labour – Topic 7 & Goldin & 10 & 0.668 & 0.593 & Topic precedes citations \\
Growth – Topic 9 & Acemoglu et al. & 9  & 0.587 & 0.214 & Topic clearly precedes citations \\
Growth – Topic 8 & Banerjee, Duflo \& Kremer & 0  & 0.601 & 0.601 & Contemporaneous \\
Finance – Topic 4 & Acemoglu et al. & 10 & -0.636 & -0.462 & Weak / misaligned \\
\bottomrule
\end{tabular}
\begin{tablenotes}
\footnotesize
\item Positive lags indicate that increases in topic prevalence precede citation growth. Negative lags indicate the reverse.
\end{tablenotes}
\end{threeparttable}
\end{sidewaystable*}

\section{Discussion and Conclusion}
\label{section5}
\paragraph{}
This article has developed a statistical framework for identifying narrative emergence in longitudinal text corpora. By defining emergence as a sustained increase in aggregated topic proportions estimated via LDA, the framework translates a qualitative theoretical concept into an empirically testable statistical quantity.
\paragraph{}
First, the framework clarifies what constitutes evidence of emergence. Topic prevalence is a latent, model-estimated quantity. Interpreting changes in $\bar{\theta}_{k,t}$ requires explicit statistical testing rather than visual inspection. The combination of Mann-Kendall trend detection and Sen's slope estimation provides a method for distinguishing persistent structural shifts from short-term fluctuation.
\paragraph{}
Second, the empirical application demonstrates that influential research strands are detectable when they are expressed through concentrated thematic vocabularies. Financial intermediation, institutional political economy, experimental development economics, and gender and labour markets emerge as lexically coherent topics with sustained upward trajectories. By contrast, contributions that extend existing paradigms or operate methodologically across domains may diffuse within broader thematic structures rather than forming distinct clusters. Narrative detectability therefore depends, in this context, not only on influence, but on lexical concentration.
\paragraph{}
Third, the citation alignment analysis shows that narrative emergence often coincides with, or precedes, formal markers of disciplinary recognition. In some subfields, thematic prominence and citation growth evolve contemporaneously; in others, topic prevalence leads citation accumulation by nearly a decade. This heterogeneity highlights the value of combining internal model-based diagnostics with external validation metrics.
\paragraph{}
The framework also has limitations. LDA assumes a fixed number of topics and exchangeability at the document level. Estimated topic proportions depend on preprocessing choices, prior settings, and the selected number of topics. Citation counts capture only one dimension of influence and may lag behind broader intellectual change. Future research could extend the framework to examine how exogenous events influence the emergence and evolution of narratives. External shocks such as major publications, policy interventions, or economic crises may induce abrupt reallocations of thematic attention, and modelling such interventions explicitly could help distinguish endogenous narrative growth from externally triggered shifts, for example by incorporating approaches that treat variation in textual communication itself as a source of shocks \citep{handlan2020text}.
\paragraph{}
More broadly, the contribution lies in demonstrating that narrative emergence can be operationalised as a statistically tractable phenomenon. Rather than treating narratives as purely interpretative constructs, the framework shows that their rise can be measured as a persistent reallocation of thematic attention within discourse. When combined with external indicators, topic trajectories provide a reproducible lens for studying intellectual transformation over time.
\paragraph{}
While illustrated using economics journals, the methodology is applicable to any longitudinal corpus in which thematic change unfolds gradually — including scientific fields, policy debates, and media discourse. In this sense, the framework contributes to bridging narrative theory and statistical text analysis, offering a principled basis for analysing how ideas rise, diffuse, and consolidate within structured discursive systems.
%==========================
% Session information
%==========================

% Session number (mandatory)
% Example: \def\sessionnumber{CS123}
%\def\sessionnumber{CS000}

% Session name (mandatory)
% Example: \def\sessionname{Probability and Statistics}
%\def\sessionname{Contributed Session Name}

% First organizer (mandatory)
% Example: \def\firstorganizer{Name and Surname}
%\def\firstorganizer{Name Surname}

% Second organizer (optional)
% Leave empty or remove this line if there is only one organizer
% Example:
% \def\secondorganizer{}
%\def\secondorganizer{Name Surname}

%==========================

% Do NOT edit the following part
%==========================

% This command defines how the second organizer is printed.
% If \secondorganizer is empty, nothing is printed.
% Otherwise, the output is: "and <Second Organizer>".
%\newcommand{\andsecondorganizer}{%
%  \ifx\secondorganizer\empty
 % \else
 %   \textnormal{ and \secondorganizer}%
%  \fi
%}

%==========================
% Session header (output)
%==========================

% Print the contributed session information.
% The command \andsecondorganizer automatically includes the second organizer
% only if it is defined.
%{\footnotesize
%\noindent\textbf{Contributed Session}\\
%\textbf{\sessionnumber}: \sessionname\ organized by 
%\firstorganizer\andsecondorganizer}
%
%
%
%  ---- Bibliography ----
%
% If you use BibTeX, specify bibliography style 'splncs04'.
% This will format the references in the correct Springer LNCS style.
%
% \bibliographystyle{splncs04}

\section*{Acknowledgements}
The authors thank Kevin Wilson for helpful suggestions during the revision of this work, particularly regarding Section \ref{sec:citations}. Any errors are our own.

\bibliographystyle{apalike}
\bibliography{name}

\end{document}